# iPrivacy: A Distributed Approach to Privacy on the Cloud

Ernesto Damiani, Francesco Pagano
Department of Information Technology
Università degli Studi di Milano
Milano, Italy
{ernesto.damiani, francesco.pagano}@unimi.it

Davide Pagano
School of Engineering
Politecnico di Milano
Milano, Italy
davide1.pagano@mail.polimi.it

*Abstract*—The increasing adoption of Cloud storage poses a number of privacy issues. Users wish to preserve full control over their sensitive data and cannot accept that it is accessible by the remote storage provider. Previous research was made on techniques to protect data stored on untrusted servers; however we argue that the cloud architecture presents a number of open issues. To handle them, we present an approach where confidential data is stored in a highly distributed database, partly located on the cloud and partly on the clients. Data is shared in a secure manner using a simple grant-and-revoke permission of shared data and we have developed a system test implementation, using an in-memory Relational Data Base Management System with row-level data encryption for fine-grained data access control.

I. INTRODUCTION

Cloud computing is the commercial evolution of grid computing [23]; it provides users with readily available, pay-as-you-go computing and storage power, allowing them to dynamically adapt their IT (Information Technology) costs to their needs. In this fashion, users need neither costly competence in IT system management nor huge investments in the start-up phase in preparation for future growth.

While the cloud computing concept is drawing much interest, several obstacles remain to its widespread adoption, including:
- Current limits of ICT infrastructure: availability, reliability and quality of service;
- Different paradigm of development of cloud applications with respect to those used for desktop applications;
- Privacy risks for confidential information residing in the cloud.

Hopefully, the first obstacle will diminish over time, thanks to increasing network availability; the second will progressively disappear by training new developers; the third issue however, is still far from being solved and may impair very seriously the real prospects of cloud computing.

In this paper, we illustrate some techniques for providing data protection and confidentiality in outsourced databases (Section II), analyze some possible pitfalls of these techniques in Cloud Computing (Section III), and propose a new solution based on distributed systems (Section IV), experimentally implemented and benchmarked (Section V).

I. THE PROBLEM OF PRIVACY

The cloud infrastructure can be accessible to public users (Public Cloud) or only to those operating within an organization (Private Cloud) [3]. Generally speaking, external access to shared data held by the cloud goes through the usual authentication, authorization, and communication phases. The access control problem is well acknowledged in the database literature and available solutions guarantee a high degree of assurance.

However, the requirement that the maintainer of the datastore cannot access or alter outsourced data is not easily met, especially on public clouds like Google App Engine for Business, Microsoft Azure Platform or Amazon EC2 platform.

Indeed, existing techniques for managing the outsourcing of data on untrusted database servers [13] [14] cannot be straightforwardly applied to public clouds, due to several reasons:
- The physical structure of the cloud is, by definition, undetectable from the outside; who is really holding the data stored on the cloud?
- The user often has no control over data replication; i.e., how many copies exist (including backups) and how are they managed?
- The lack of information on the geographical location of data (or its variation over time) may lead to jurisdiction conflicts when different national laws apply.

In the next section, we will briefly summarize available techniques for data protection on untrusted servers, and show how their relation to the problems outlined above.

*A. Data Protection*

To ensure data protection in outsourcing, the literature reports three major techniques [6]:
- Data encryption [15];
- Data fragmentation and encryption [16], which in turn can be classified into two major techniques:
  - non-communicating servers [17][18];
  - unlinkable fragments [19];
- Data fragmentation with owner involvement [20].

*1) Data encryption*

To prevent unauthorized access by the Datastore Manager (DM) of the outsourced Relational Data Base Management Systems (RDBMS), data is stored in encrypted form. Obviously, the DM does not know the encryption keys, which are stored apart from the data. The RDBMS receives an encrypted database and works on bit-streams that only the clients, who hold the decryption keys, can interpret correctly.

Figure 1 shows the transformation of a plain text tuple into an encrypted one.

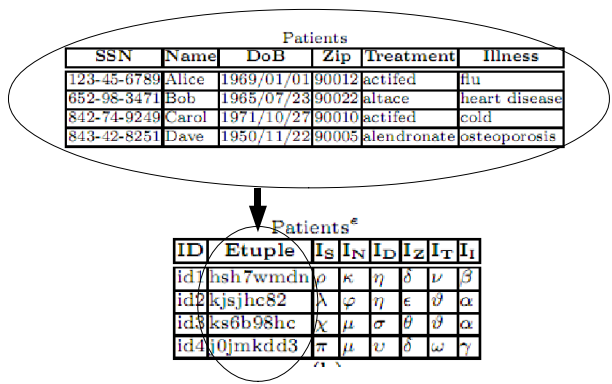

Figure 1. Data encryption, source: [6]

Decryption keys are generated and distributed to trusted clients by the data owner or by a trusted delegate.

Encryption can be performed at different levels of granularity: field, record, table, db [28]. Usually, the level adopted is the record (i.e., a tuple in relational databases).

It is important to remark that since data is encrypted, the DBMS cannot index it based on plaintext and therefore it cannot resolve all queries. Available proposals solve this problem by providing, for each encrypted field to be indexed, an additional indexable field, obtained by applying a non-injective transformation $f$ to plaintext values (e.g., a hash of the field's content). Using this method, equality queries can be performed easily, although with a precision index < 1 (to prevent statistical data mining). The trusted client, after receiving the encrypted result set for the query, will decrypt it and exclude spurious tuples. However range queries are difficult to compute, since the transformation $f$ in general will not (and should not) preserve the order relations of the original plaintext data. Specifically, it will be impossible for the outsourced RDBMS to answer range queries that cannot be reduced to multiple equality conditions (e.g., $1<=x<=3$ can be translated into $x=1$ or $x=2$ or $x=3$) unless specific techniques are applied. In literature, there are several proposals for $f$, including:

1. *Domain partitioning* [24]*:* the domain is partitioned into equivalence classes, each corresponding to a single value in the codomain of $f$;

2. *Secure hashing* [13]: secure one-way hash function, which takes as input the clear values of an attribute and returns the corresponding index values. $f$ must be deterministic and non-injective.

To handle range queries, a solution, among others, is to use an encrypted version of a B ± tree to store plaintext values, and maintain the values order. Because the values have to be encrypted, the tree is managed at the Client side and it is read-only in the Server side. Alternatively, the position information of each field in the original relation can be added to the encrypted data [33].

Let us, now, consider data protection strategies based on partitioning.

*1) Data fragmentation*

Normally, of all the outsourced data, only some columns and/or some relations are confidential, so it is possible to split the outsourced information in two parts, one for confidential and one for public data. Its aim is to minimize the computational load of encryption/decryption.

*a) Non-communicating servers*

In this technique, two *split databases* are stored, each in a different untrusted server (called, say, $S_1$ and $S_2$). The two untrusted servers need to be independent and non-communicating to prevent their alliance and reconstruction of the complete information. In this situation, the information may be stored as plaintext at each server.

Each Client query is decomposed in two subqueries: one for $S_1$ and one for $S_2$. The result sets have to be related and filtered, by the Client.

Figure 2 schematizes the resulting structure.

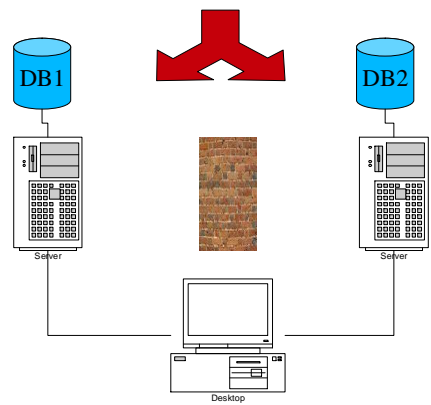

Figure 2. Non-communicating servers

*b) Unlinkable fragments*

In reality, it is not easy to ensure that split servers do not communicate; therefore the previous technique may be inapplicable. A possible remedy is to divide information in two or more fragments. Each fragment contains all the fields of the original information, but some are in clear form while the others are encrypted. To protect encrypted values from frequency attacks, a suitable *salt* is applied to each encryption. Fragments are guaranteed to be unlinkable (i.e., it is impossible to reconstruct the original relation and to determine the sensitive values and associations without the decrypting key). These fragments may be stored in one or more servers.

Each query is then decomposed in two subqueries:
- The first, executed on the Server, chooses a fragment (all fragments contain the entire information) and selects tuples from it according to clear text values.

- It returns a result set where some fields are encrypted;
- The second, executed on the Client (only if encrypted fields are involved in the query), decrypts the information and removes the spurious tuples.

Figure 3 schematizes the resulting structure.

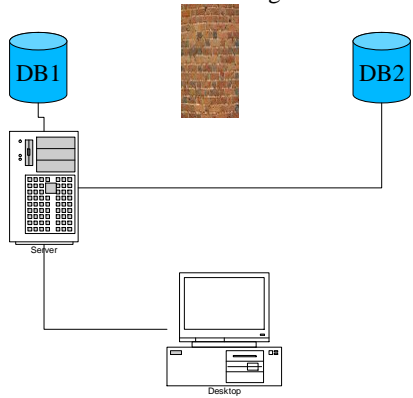

Figure 3. Unlinkable fragments

*2) Data fragmentation with owner involvement*

Another adaptation of the *non-communicating servers* technique consists of storing locally the sensitive data and relations, while outsourcing storage of the generic data. So, each tuple is split in a server part and in a local part, with the primary key in common. The query is then resolved as shown above.

Figure 4 schematizes the resulting structure.

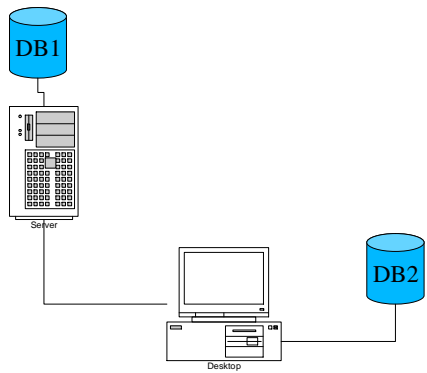

Figure 4. Data fragmentation with owner involvement

### B. Selective access

In many scenarios, access to data is selective, with different users enjoying different views over the data. Access control can discriminate between read and write operations on an entire record or only on a part of it.

An intuitive way to handle this issue is to encrypt different portions of data with different keys that are then distributed to users according to their access privileges. To minimize overhead it is required that:

- No more than one key is released to each user;
- Each resource is encrypted not more than once.

To achieve these objectives, a hierarchical organization of keys can be envisioned. Basically, users with the same access privileges are grouped and each resource is encrypted with a key corresponding to the set of users that can access it. This way, a single key can be possibly used to encrypt more than one resource.

*1) Dynamic rights management*

Should the user's rights change over time (e.g., the user changes department) it is necessary to remove that user from a group/role as follows:

- Encrypt data by a new key;
- Remove the original encrypted data;
- Send the new key to the rest of the group.

Note that these operations must be performed by data owner because the untrusted DBMS has no access to the keys. This active role of the data owner goes somewhat against the reasons for choosing to outsource data in the first place.

*a) Temporal key management*

An important issue, common to many access control policies, concerns time-dependent constraints of access permissions. In many real situations, it is likely that a user may be assigned a certain role or class for only a limited time. In such case, users need a different key for each time period. A time-bound hierarchical key assignment scheme is a method to assign time-dependent encryption keys and private information to each class in the hierarchy in such a way that key derivation depends also on temporal constraints. Once a time period expires, users in that class should not be able to access any subsequent keys unless further authorized to do so [9].

*b) Database replica*

In [7], the authors, exploiting the never ending trend to a lower price-per-byte in storage, propose to replicate $n$ times the source database, where $n$ is the number of different roles having access to the database. Each database replica is a view, entirely encrypted using the key created for the corresponding role. Each time a role is created, the corresponding view is generated and encrypted with a new key expressly generated for the newly created role. Users do not own the real keys, but receive a token that allows them to address a request-to-cipher to a set KS of key servers on the cloud.

### C. A document base sample: Crypstore

Crypstore is a non-transactional architecture for the distribution of confidential data, whose structure is shown in Figure 5. Crypstore's Storage Server contains data in encrypted form, so it cannot read them. User who wants to access data is authenticated at the Key Servers with the certificate issued by the Data Administrator and requires the decryption key. The Key Servers are $N$ and, to ensure that none of them knows the whole decryption key, each of them contains only a part of the encryption key. To rebuild the key, only $M$ ($<N$) parts of key are needed; redundancy provides greater robustness to failures and attacks (e.g., Denial of Service attacks).

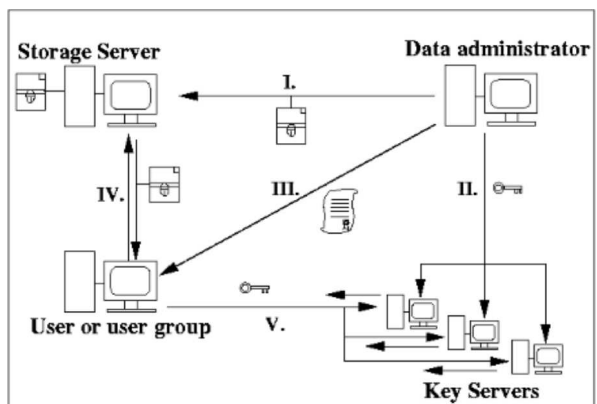

Figure 5. Crypstore

Really, Crypstore is an application of the time-honored "divide and conquer" technique, where data is separated from decryption keys.

Here privacy is not entirely guaranteed because, theoretically at least, the owner of Key Servers and the Storage Server may collude. The only way to exclude this (however remote) possibility is to have trusted Key Servers, but this would be equivalent to store the data directly, as plaintext, in a trusted storage. In practice, however, the probability of collusion decreases with the number of players involved and can be safely ignored in many cases.

## II. PRIVACY WITHIN THE CLOUD

All techniques discussed above are based on data encryption and/or data fragmentation using full separation of roles and of execution environments between the user and the datastore (and possibly the keystore) used to manage the outsourced data.

Let us now compare the assumptions behind such techniques with two of the basic tenets of current cloud computing architectures: data and applications being on the "same side of the wall", and data being managed via semantic datastores rather than by a conventional RDBMS.

### A. On the same side of the wall

*Ubiquitous access* is a major feature of cloud computing architectures. It guarantees that cloud application users will be unrestrained by their physical location (with internet access) and unrestrained by the physical device they use to access the cloud.

To satisfy the above requirements (in particular the second), we normally use thin clients, which run cloud applications remotely via a web user interface.

The three main suppliers of Public Cloud Infrastructure (Google App Engine for Business, Amazon Elastic Compute Cloud and Windows Azure Platform) all include a datastore, and an environment for remote execution summarized in Tables I and II:

TABLE I. DATASTORE SOLUTIONS USED BY PUBLIC CLOUDS

| Environment | Datastore |
|---|---|
| Google | Bigtable |
| Amazon | IBM DB2 |
| | IBM Informix Dynamic Server |
| | Microsoft SQLServer Standard 2005 |
| | MySQL Enterprise |
| | Oracle Database 11g |
| | Others installed by users |
| Microsoft | Microsoft SQL Azure |

TABLE II. EXECUTION ENVIRONMENTS USED BY PUBLIC CLOUDS

| Environment | Execution environment |
|---|---|
| Google | J2EE (Tomcat + GWT) |
| | Python |
| Amazon | J2EE (IBM WAS, Oracle WebLogic Server) and others installed by users |
| Microsoft | .Net |

In all practical scenarios, public cloud suppliers handle both data and application management.

If the cloud supplier is untrustworthy, it can intercept communications, modify executable software components (e.g., using aspect programming), monitor the user application memory, etc.

Hence, available techniques for safely outsourcing data to untrusted DBMS no longer guarantee the confidentiality of data outsourced to the cloud.

The essential point consists in having the data and the user interface application logic *on the same side of the wall* (see Figure 6).

This is a major difference w.r.t. outsourced database scenarios, where presentation was handled by trusted clients. In the end, the data must be presented to the user in an intelligible and clear form; that is the moment when a malicious agent operating in the cloud has more opportunities to intercept the data. To prevent unwanted access to the data at presentation time, it would be appropriate moving the presentation logics off the cloud to a trusted environment that may be an intranet or, at the bottom level, a personal computer.

However, separating data (which would stay in the cloud) from the presentation logics may enable the creation of local copies of data, and lead to an inefficient cooperation between the two parts.

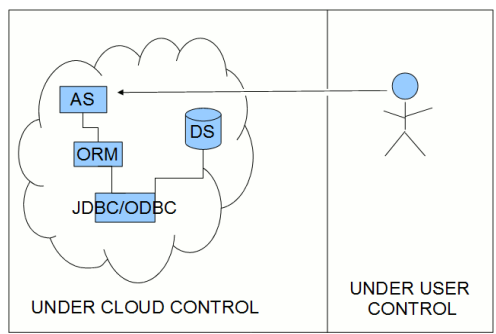

Figure 6. The wall

## B. Semantic datastore

Cloud computing solutions largely rely on semantic (non-relational) DBMS. These systems do not store data in tabular format, but following the natural structure of objects. After more than twenty years of experimentation (see, for instance, [10] for the Galileo system developed at the University of Pisa), today, the lower performance of these systems is no longer a problem. In the field of cloud computing, there is a particular attention to Google Bigtable.

"*Bigtable is a distributed storage system for managing structured data that is designed to scale to a very large size: petabytes of data across thousands of commodity servers. In many ways, Bigtable resembles a database: it shares many implementation strategies with databases.*" [11]

With a semantic datastore like Bigtable, there is a more strict integration between in-memory data and stored-data; they are almost indistinguishable from the programmer viewpoint. There are not distinct phases when the program loads data from disk into main memory or, in the opposite direction, when program serialize data on disk. Applications do not even know where the data is stored, as it is scattered over the cloud.

In such a situation, the data outsourcing techniques discussed before cannot be applied directly, because they were designed for untrusted RDBMS.

## III. OUR APPROACH

We are now ready to discuss our new approach to the issue of cloud data privacy. We build over the notion introduced in [7] of defining a view for every user group/role, but we prevent performance degradation by keeping all data views in the user environment.

Specifically, we atomize the application/database pair, providing a copy per user. Every instance runs locally, and maintains only authorized data that is replicated and synchronized among all authorized users.

In the following subsections we will analyze our solution in detail.

### A. Information sharing by distributed system

We will consider a system composed of:
1. Local agents distributed at client side;
2. A central synchronization point.

Figure 7 shows the proposed architecture:

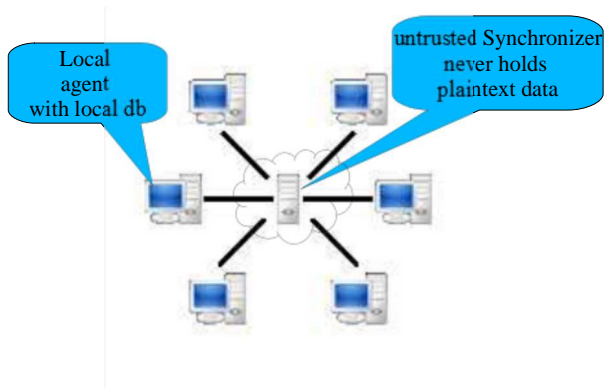

Figure 7. The architecture

### 1) The model

Henceforth, we will use the term *dossier* to indicate a set of related information. Our data model may be informally represented by the diagram in Figure 8.

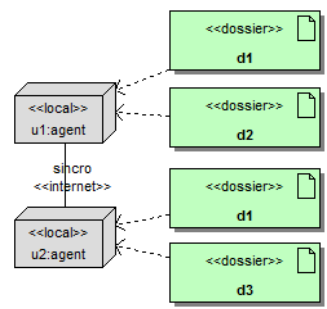

Figure 8. The model

In our model, each node represents a local, single-user application/database dedicated to an individual user ($u_n$). The node stores only the dossiers that $u_n$ owns. Shared dossiers (in this example, $d_1$) are replicated on each node. When a node modifies a shared dossier, it must synchronize, also using heuristics and learning algorithms, with the other nodes that hold a copy of it. Below we give a simple SWOT analysis of this idea.

### 2) Strength/Opportunities
- Information sharing using untrusted Synchronizer;
- Small amount of local data, less attractive for attackers;
- Only the final user has clear-text information;
- Unrestrained individual nodes, that can also work offline (with deferred synchronization);
- Simplicity of data management (single user);
- Completeness of local information.

To clarify the last point, suppose that the user $u_n$ wants to know the number of the dossier she is treating. In a classic intranet solution, where dossiers reside on their owners' servers, in addition to its database, $u_n$ should examine the data stores of all other collaborating users. With our solution, instead, $u_n$ will simply perform a local query because the dossiers are replicated at each client.

*3) Weaknesses/Threats*
- Complexity of deferred synchronization schemes [21];
- Necessity to implement a mechanism for grant/revoke and access control permissions.

This last point is particularly important and it deserves further discussion:
- Each user (except the data owner) may have partial access to a dossier. Therefore each node contains only the allowed portion of the information;
- Authorization, i.e., granting to a user $u_j$ access to a dossier $d_k$, can be achieved by the data owner simply by transmitting to each node only the data it is allowed to access;
- The inverse operation can be made in the case of a (partial or complete) revocation of access rights. An obvious difficulty lies in ensuring that data becomes no longer available to the revoked node. This is indeed a moot point, as it is impossible – whatever the approach - to prevent trusted users from creating local copies of data while they are authorized and continue using them after revocation. We are evaluating the opportunity to use watermarking for relational databases [26] to provide copyright protection and tamper detection.

B. *Proposed solution*

We are now ready to analyze our solution in detail. To simplify the discussion, we introduce the following assumptions:
- Each dossier has only one owner;
- Only the dossier's owner can change it.

These assumptions permit the use of an elementary cascade synchronization in which the owner will submit the changes to the receivers. However, they can be relaxed at the cost of a higher complexity in synchronization [34].

Our solution consists of two parts: a trusted client and a remote untrusted synchronizer (see Figure 9).

The client maintains local data storage where:
- The dossiers that she owns are (or at least can be) stored as plain-text;
- The others, instead, are encrypted each with a different key.

The Synchronizer stores the keys to decrypt the shared dossiers owned by the local client and the modified dossiers to synchronize.

When another client needs to decrypt a dossier, she connects to the Synchronizer and obtains the corresponding decryption key.

The data and the keys are stored in two separate entities, none of which can access information without the collaboration of the other part.

*1) Structure*

From the architectural point of view, we divide our components into two packages, a local (client agent), which contains the dossier plus additional information such as access lists, and a remote (global synchronizer), which contains the list of dossiers to synchronize, their decryption keys and the public keys of clients.

A UML view of involved classes is shown in Figure 10.

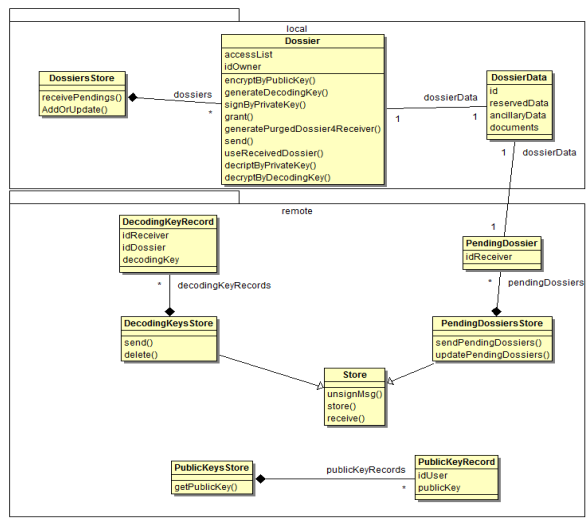

Figure 10. Class view

*2) Grant*

An owner willing to grant rights on a dossier must follow the sequence shown in Figure 11:

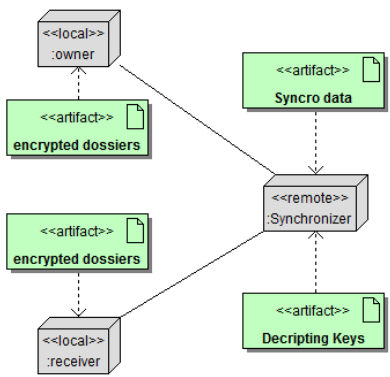

Figure 9. Deployment diagram of distributed system

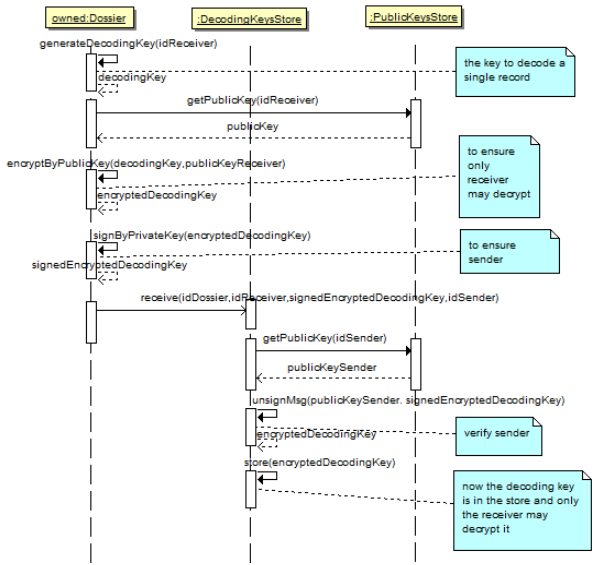

Figure 11. Grant sequence

Namely, for each receiver, the owner:
- Generates the decryption key
- Encrypts it with the public key of the receiver to ensure that others cannot read it
- Signs it with its private key to ensure its origin
- Sends it to the Synchronizer, which verifies the origin and adds it to the storage of the decoding keys. The key is still encrypted with the public key of the receiver, so only the receiver can read it.

*3) Send*

When an owner modifies a dossier, she sends it to the Synchronizer the sequence shown in Figure 12:

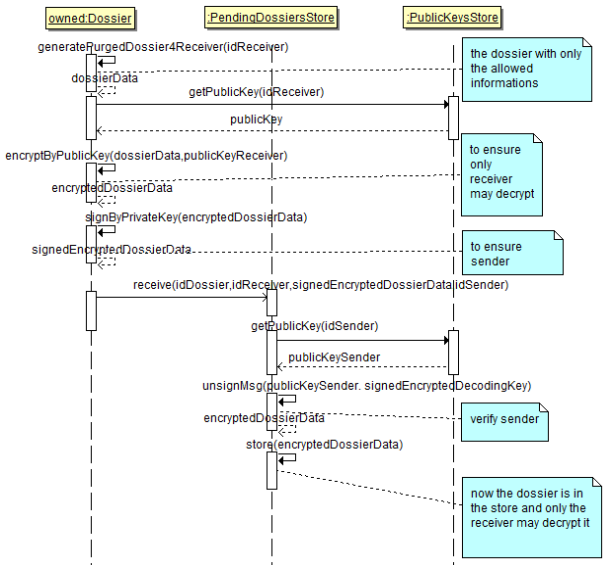

Figure 12. Send sequence

For each receiver, the owner:
- Generates a "pending dossier" by removing information that the receiver should not have access to;
- Encrypts the pending dossier with the previously generated decryption key;
- Signs with his own private key to certificate its origin;
- Sends it to the Synchronizer, which verifies the origin and adds it to the storage of "pending dossiers". Again, the dossier is still encrypted with the public key of the receiver, so only the receiver can read it.

*4) Receive*

Periodically, each client updates un-owned dossiers by following the sequence shown in Figure 13:

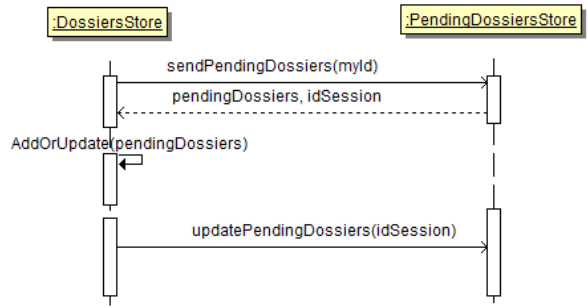

Figure 13. Receive sequence

Each client:
- Requests the "pending dossiers" to the Synchronizer.
- Stores the (still encrypted) dossier in the local storage;
- Removes the received dossiers from the Synchronizer.

*5) Use*

When a client needs to access an unowned (encrypted) dossier, the sequence shown in Figure 14 is used:

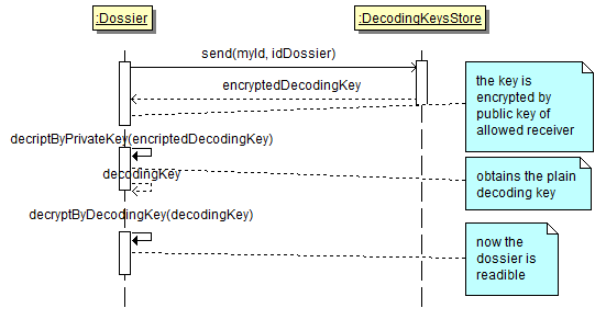

Figure 14. Use sequence

The client:
- Asks the Synchronizer for the decryption key (that is encrypted by her public key);
- Decrypts it with her private key;
- Decrypts the dossier by the resulting decryption key.

If the decryption key does not exist, two options are available:
- The record is deleted from the local datastore because a revoke happened;
- The record remains cached (encrypted) into the local datastore because access rights to it could be restored.

  *6) Revoke*

To revoke access to a receiver, it is sufficient to delete the corresponding decryption key from the Synchronizer. The sequence diagram is shown in Figure 15.

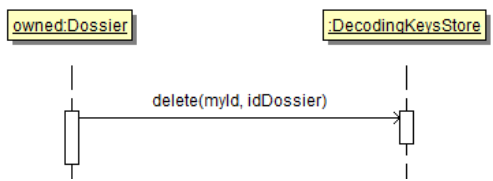

Figure 15. Revoke sequence

## IV. EXPERIMENTATION

To experiment with our architecture we implemented the custom client and Synchronizer. The client needs to use row-level encryption. In a normal RDBMS, however, this technique has significant disadvantages in terms of performance and functionality: querying would be possible only through the construction of appropriate indexes for each column of the table (with a considerable waste of resources both in terms of time and space), while the constraints and foreign keys would be almost unusable.

Another major issue concerns the management of keys: row-level encryption could potentially lead to the generation and maintenance (and / or distribution) of a key for each row of each table encrypted with this method. To solve (or reduce) the concern, we use some advanced techniques of key management, such as:
- Broadcast (or Group) encryption [32]: rows are divided into equivalence classes, based on recipients. Every class is encrypted using an asymmetric algorithm where the encryption key is made in a way that each recipient can decrypt the information using only its own private key. Both the public and the private keys are generated by a trusted entity.
- Identity Based Encryption [30]: it bounds the encryption key to the identity of the recipient. Each recipient generates by itself a key pair used to encrypt/decrypt information.
- Attribute Based Encryption [31]: it bounds the encryption key to an attribute (a group) of recipient. Each recipient receives from a trusted entity the private key used to decrypt, while the sender calculates the encryption key.

The complexity of these techniques is a major reason why conventional RDBMSs do not use encryption at the row-level.

*A. In memory databases*

An in-memory database (IMDB, also known as main memory database system or MMDB and as real-time database or RTDB) is a database management system that primarily relies on main memory for computer data storage [35]. It is important to remark that, while a conventional database system stores data on disk but caches it into memory for access, in an IMDB the data resides permanently in the main physical memory and there is a backup copy on disk [27].

In-memory databases have recently become an intriguing topic for the database industry. With the mainstream availability of 64-bit servers with many gigabytes of memory a completely RAM based database solution is a tempting prospect to a much wider audience [36].

IMDBs are intended either for personal use (because they are comparatively small w.r.t. traditional databases), or for performance-critical systems (for their very low response time and very high throughput). They use main memory structures, so they need no translation from disk to memory form, and no caching and they perform better than traditional DBMSs with Solid State Disks.

Normally, the use of volatile memory-based IMDBs supports the three ACID properties of atomicity, consistency and isolation, but lacks support for the durability property. To add this when non-volatile random access memory (NVRAM) is not available, IMDBs use a combination of transaction logging and primary database check-pointing to the system's hard disk: they log changes from committed transactions to physical medium and, periodically, update a disk image of the database. Having to write updates to disk, the write operations are heavier than read-only. Logging policies vary from product to product: some leave the choice of when to write the application on file, others do all the checkpoints at regular intervals of time or after a certain amount of data entered / edited.

In Table III, we summarize pros and cons for IMDBs.

TABLE III. IMDBs PROS AND CONS

| Pros | Cons |
|---|---|
| Fast transactions | Complexity of durability's implementation |
| No translation | |
| High reliability | Size limited by main memory |
| Multi-User Concurrency (few locks) | |

Obviously, the limitation of this type of database is related to the amount of RAM on computer hosting the db. But given their nature, IMDBs are well suited to be distributed and replicated across multiple nodes to increase capacity and performance.

The proposed solution works around this limitation: not having a single central database containing the whole data,

we preferred to give one database for each client application. This database contains only owned data, while external data will be added (or removed) via the synchronizer, based on access permissions.

To minimize cryptography overhead, we encrypt only rows "received" by other nodes, while rows owned by the local node are stored in cleartext form.

Well-known open solutions of IMDB are Apache Derby, HyperSQL (HSQLDB) and SQLite. For our implementation, we chose to use HyperSQL rel. 2.0.

*1) HyperSql*

HyperSQL [37] is a pure Java RDBMS. Its strength is, besides the lightness (about 1.3Mb for version 2.0), the capability to run either as a Server instance either as a module internal to an application (in-process).

A database started "in-process" has the advantage of speed, but it is dedicated only to the containing application (no other application can query the database). For our purposes, we chose server mode. In this way, the database engine runs inside a JVM and will start one or more "in-process" databases, listening requests from processes in the local machine or remote computers.

For interactions between clients and database server, we can use three different protocols:
- HSQL Server: the fastest and most used. It implements a proprietary communication protocol;
- HTTP Server: it is used when access to the server is limited only to HTTP. It consists of a web server that allows JDBC clients to connect over http;
- HTTP Servlet: as the Http Server, but it is used when accessing the database is managed by a servlet container or by an application servlet (e.g., Tomcat). It is limited to using a single database.

Several different types of databases (called catalogs) can be created with HyperSQL. The difference between them is the methodology adopted for data storage:
- Res: this type of catalog provides for the storage of data into small JAR or ZIP files;
- Mem: data is stored completely in the machine's RAM, so there is no persistence of information outside of the application life cycle in the JVM;
- File: data is stored in files residing into the file system of the machine.

In our work, we used the latter type of databases.

A catalog file can use up to six files on the file system for its operations. The name of these files consists of the name of the database plus a dot suffix.

Assuming we have a database called "db_test", files will be:
- db_test.properties containing the basic settings of the DB;
- db_test.log: used to periodically save data from the database, to prevent data loss in case of a crash;
- db_test.script: containing the table definitions and other components of the DB, plus data of not-cached tables;
- db_test.data: containing the actual data of cached tables. It can be not present in some catalogs;
- db_test.backup: containing the compressed backup of last ".data" file, that may be not present in some catalogs;
- db_test.lobs: used for storing BLOB or CLOB fields.

Besides these files, HyperSQL can connect to CSV files.

A client application can connect to HyperSQL server using the JDBC driver (.Net and ODBC drivers are "in late stages of development"), specifying the type of database to access (file, mem or res).

HyperSQL implements the SQL standard either for temporary tables either for persistent ones. Temporary tables (TEMP) are not stored on the file system and their life cycle is limited to the duration of the connection (i.e., of the Connection object). The visibility of data in a TEMP table is limited to the context of connection used to populate it. With regard to the persistent tables, instead, HyperSQL provides three different types of tables, according to the method used to store the data:
- MEMORY: it is the default option when a table is created without specifying the type. *Memory table* data is kept entirely in memory, while any change to its structure or contents is recorded in .log and .script files. These two files are read at the opening of database to load data into memory. All changes are saved when closing the database. These processes can take a long time in the case of tables larger than 10 MB.
- CACHED: when this type of table is chosen, only part of the data (and related indexes) is stored in memory, thus allowing the use of large tables at the expense of performance.
- TEXT: the data is stored in formatted files such as .csv.

In our implementation, we use MEMORY tables.

The Loader and the Serializer are the main parts of HyperSQL that we analyzed and modified. They are the mechanisms that load the data from text files at the opening and save them to the database at closing.

B. *Implemented solution*

*1) Client side*

On the client side, using IMDBs, we have only two interactions between each local agent and the Synchronizer, as shown in Figure 16.

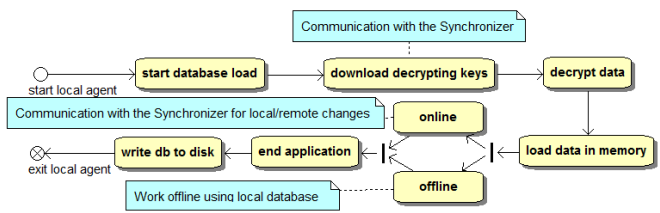

Figure 16. State diagram of client

We have modified the classes included in file hsqldb.jar to handle encryption. The basic idea was to manage encryption in the .log and .script text files. The rows that are owned by the local client are stored in clear-text, while the shared rows "granted" by other owners are stored encrypted.

The values contained in tables are stored in form of SQL insert:

INSERT INTO table_name(field_1, field_2, ..., field_n) VALUES(value_1, value_2, ..., value_n)

Earlier, to obtain control access granularity at the field level, we encrypted field by field. This way, the text contained in the database file is in the form of:

INSERT INTO table_name(field_1, field_2, ..., field_n) VALUES(pk, encrypted_value_2, ..., encrypted_value_n)

The primary key *pk* needs to be in clear-text, since it is used to retrieve the decrypting keys from the central Synchronizer. We dropped this idea because it requires changing the I/O code for each possible database type and an attacker may obtain some information such as table, primary key and number of rows.

Our current solution is to encrypt the whole row by AES symmetric algorithm. The encryption overhead is lower than the previous solution and all information is hidden to curious eyes. To relate the encrypted row (stored locally) to the decrypting key (stored in the remote Synchronizer), we use a new key (*id_pending_row*). The encrypted row is prefixed by a clear-text header containing the id_pending_row delimited by "$" and "@". The encrypted value is then stored in a hexadecimal representation, so a generic row is of the form:

$27@5DAAAED5DA06A8014BFF305A93C957D

*a) Load time*

At load time, the .script file will contain clear-text and encrypted rows, e.g.:

INSERT INTO students(id,name) VALUES(12,'Alice');
INSERT INTO students(id,name) VALUES(31,'Bob');
$27@5F3C25EE5738DAAAED5DA06A80F305A93C95A
$45@5DA67ADA06AAED580FA914BF3C953057D387F
INSERT INTO students(id,name) VALUES(23,'Carol');

The class whose task is reading the file and loading the appropriate data in memory is *ScriptReaderText*.

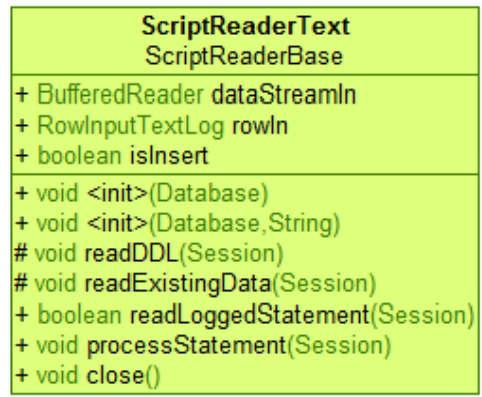

Figure 17. UML of ScriptReaderText class

The *readLoggedStatement* method parses each line of text in the .log or .script files and forwards the result to the *processStatement* method, which loads data into memory.

We changed the *readLoggedStatement* method to make a preprocessing: if it finds a record header (enclosed between $ and @) in the text line, it extracts the *id_pending_row_received*. Using this id, the client requests to the central Synchronizer the related decoding key, which it uses to decrypt the entire text line and to proceed with normal HyperSQL management. If the decoding key is unavailable, the text line is temporarily discarded (it is not deleted if it was not received for communication problem with the Synchronizer).

*b) Save time*

The class ScriptWriterText manages the write operations in .log and .script files.

The affected methods are *writeRow* and *writeRowOutToFile*.

The former deals with building the string that will be written into the text file (INSERT INTO ....) wich corresponds to the in memory data. A *Table* instance contains the information about the table structure (table name, field names, types of data, constraints, etc.). The values of fields are in an array of *Object*. The SQL *insert* is written in a text buffer that is stored in the .script file by the method *writeRowOutToFile*. Because each table has an *id_pending_row_received* column, we modified the *writeRow* method to check if the row is owned or shared by another user. In the latter case (*id_pending_row_received* not null), the custom *writeRowOutToFileCrypto* method is used instead of the *writeRowOutToFile* method. *WriteRowOutToFileCrypto* uses the parameter *id_pending_row_received* to query the related symmetric encryption key from the Synchronizer, needed to encrypt the whole buffer. The result is a hexadecimal sequence, which is prefixed by the below header with id_pending_row_received.

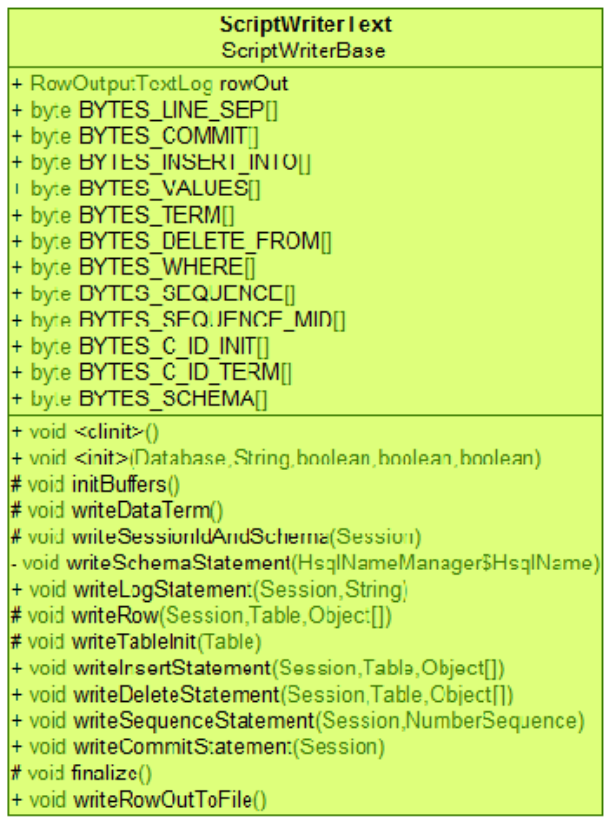

Figure 18. UML of ScriptWriterText class

*2) Server side*

When a data owner adds or updates a row in the local database, it needs to distribute this change to all the related users. To do this, we put the cloud a central Synchronizer server that acts as a mailbox.

It uses a simple database with the following tables:
- Users: containing, among others, the id and public key of each user;
- Pending Rows: it contains the rows that are added/modified in the local database of the owner, until they are delivered to destination. A unique row_id is automatically assigned to each pending row. Other information is submission date, sender and receiver. The changed row is stored in encrypted form in field encrypted_row;
- Decrypting keys: contains the keys that are used to decrypt the pending rows. Other information is: sender, receiver, expiry date, id_row.

At modification time, the owner (client side) has to:
- Serialize the row;
- Generate a symmetric key to encrypt it;
- Encrypt the row;
- Encrypt the key by the public keys of receivers;
- Send the encrypted row and the decoding keys to receiver.

Because we store the serialized row, we haven't to worry about columns data types.

The Synchronizer uses RMI to expose its services to clients. The services are grouped in three interfaces:
- KeyInterface with methods related to encryption keys: depositKey, deleteDecryptingKey, getDecriptingKeyByIdPendingRow, getPublicKeyByUser;
- SynInterface with methods for sharing the rows: sendRow, getPendingRowForUser, getAllUsers, resendRow;
- RegistrationInterface to register and manage users: registerUser, SelectUserById, selectUserByIdAndPassword.

*C. Performances*

*1) Read operations*

The system uses decryption <u>only</u> at start time, when records are loaded from the disk into the main memory. Each row is decrypted none (if it is owned by local node) or just once (if it is owned by a remote node), so this is optimal for read operations. Each decryption implies an access to the remote Synchronizer to download the related decrypting key and, eventually, the modified row.

*2) Write operations*

Write operations occur when a record is inserted / updated into the db. There is no overload until the client, when online, explicitly synchronizes data with the central server. At this moment, for each modified record, the client need to:
- Generate a new (symmetric) key
- Encrypt the record
- Dispatch the encrypted data and the decrypting key to the remote synchronizer

*3) Benchmark*

We wrote a test application that uses our modified HyperSQL driver and interacts with the other clients through our Synchronizer. It has these distinct activities:
- Creation of database and sample tables
- Population of tables with sample values
- Sharing of a portion of data with another user
- Receipt of shared dossiers from other users
- Opening of the newly created (and populated) database

The application receives three parameters:
- Number of dossiers
- Number of clients involved in sharing
- Percentage of shared dossiers

To minimize communication delay, the central Synchronizer and the clients ran on the same computer. For testing purpose, it was sufficient to use only two clients (to enable data sharing). The tests used a number of dossiers varying from 1,000 to 500,000. We tested the system with 20% and 40% of shared dossiers.

The application was compared with an equivalent one with the following differences:

- It used the unmodified HyperSQL driver
- It did not share data with other clients
- When populating the database, it created the same number of dossiers than the previous application; but, after benchmarking, it added the number of shared dossiers to have the same final number of dossiers.

We benchmarked the system using single-table dossiers of about 200 bytes, in two batteries of tests; the first with 20%, and the second with 40% of shared dossiers, which numbered from 1,000 to 500,000. The results are represented by the graphs in Figures 19-21. It is worth noting that the overhead percentage of the modified solution rapidly decreases (with 100,000 dossiers it is around 10%), either in the first battery of tests (Figure 19), and either in the second (Figure 20). In the tests, the total delay (load + create + populate + receive) stay linear in the number of dossiers and is limited, even with a huge number of dossiers (Figure 21). Local results can be slightly altered by external events not preventable (e.g., garbage collector).

*D. Results*

The delay of the system is tightly bound to communications effort with the central Synchronizer. Computing overhead is limited to just one encryption per record at write time and no more than one decryption per record at read time. Since I use symmetric encryption, these operations are very fast. The benchmark demonstrates that the delay is substantially concentrated in database opening, while the subsequent use does not involve additional delays, compared to the unmodified version.

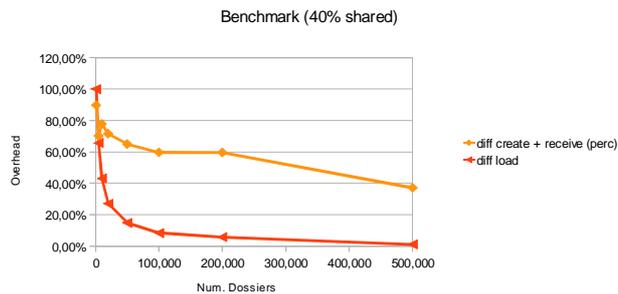

Figure 20. Overhead when 40% of dossiers are shared

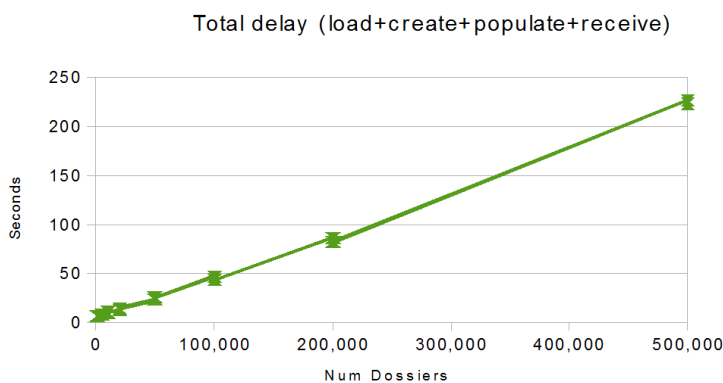

Figure 21. Total delay

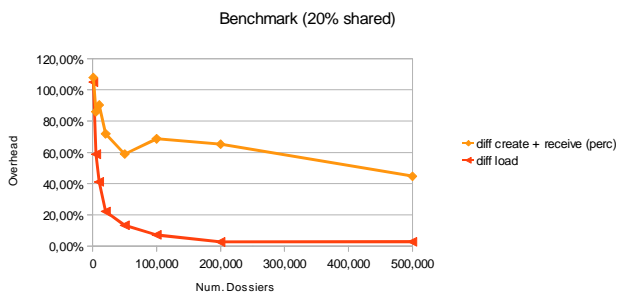

Figure 19. Overhead when 20% of dossiers are shared

V. CONCLUSIONS AND OUTLOOK

In this paper, we discussed the applicability of outsourced DBMS solutions to the cloud and provided the outline of a simple yet complete solution for managing confidential data in public clouds.

We are fully aware that a number of problems remain to be solved. A major weakness of any data outsourcing scheme is the creation of local copies of data after it has been decrypted. If a malicious client decrypts data and then it stores the resulting plain-text data in a private location, the protection is broken, as the client will be available to access its local copy after being revoked. In [22], obfuscated web presentation logic is introduced to prevent client from harvesting data. This technique, however, exposes plaintext data to cloud provider. The plain-text data manager is always the weak link in the chain and any solution must choose whether to trust the client-side or the server-side. A better solution [26] is to watermark the local database to provide tamper detection.

Another issue concerns the degree of trustworthiness of the participants. Indeed, untrusted Synchronizer never holds plain-text data; therefore it does not introduce an additional Trusted Third Party (TTP) with respect to the solutions described at the beginning of the paper. However, we need to trust the Synchronizer to execute correctly the protocols explained in this paper. This is a determining factor that our technique shares with competing solutions and, although an interesting topic, it lies beyond the scope of this paper.

In experiment phase, we introduced a simple solution to row-level encryption of databases using IMDBs. It can be used in the cloud to manage very granular access rights in a highly distributed database. This allows for stronger confidence in the privacy of shared sensitive data.

An interesting field of application is the use in (business) cooperative environments, e.g., professional networks. In these environments, privacy is a priority, but low computing resources don't allow the use of slow and complex algorithms. IMDBs and our smart encryption, instead, achieve the goal in a more effective way.

## REFERENCES

[1] E. Damiani and F. Pagano, "Handling confidential data on the untrusted cloud: an agent-based approach," Cloud Computing 2010, pp. 61-67. Lisbon, 2009. IARIA.

[2] D. Pagano and F. Pagano, "Using in-memory encrypted databases on the cloud," in press

[3] M. Armbrust, A. Fox, R. Griffith, Anthony D. Joseph, Randy H. Katz, Andy Konwinski, Gunho Lee, David A. Patterson, Ariel Rabkin, Ion Stoica, and Matei Zaharia: "A view of cloud computing", Commun. ACM 53(4), pp. 50-58 (2010)

[4] C. Jackson, D. Boneh, and J.C. Mitchell: "Protecting Browser State from Web Privacy Attacks", 15th International World Wide Web Conference (WWW 2006), Edinburgh, May, 2006.

[5] Philip A. Bernstein, Fausto Giunchiglia, Anastasios Kementsietsidis, John Mylopoulos, Luciano Serafini, and Ilya Zaihrayeu: "Data Management for Peer-to-Peer Computing : A Vision", WebDB 2002, pp. 89-94

[6] Pierangela Samarati and Sabrina De Capitani di Vimercati: "Data protection in outsourcing scenarios: issues and directions", ASIACCS 2010, pp. 1-14

[7] Nadia Bennani, Ernesto Damiani, and Stelvio Cimato: "Toward cloud-based key management for outsourced databases", SAPSE 2010, draft

[8] Mikhail J. Atallah, Marina Blanton, and Keith B. Frikken: "Incorporating Temporal Capabilities in Existing Key Management Schemes", ESORICS 2007, pp. 515-530

[9] Alfredo De Santis, Anna Lisa Ferrara, and Barbara Masucci: "New constructions for provably-secure time-bound hierarchical key assignment schemes", Theor. Comput. Sci. 407, pp.213-230 (2008)

[10] Antonio Albano, Giorgio Ghelli, M. Eugenia Occhiuto, and Renzo Orsini: "Object-Oriented Galileo", On Object-Oriented Database System 1991, pp. 87-104

[11] Fay Chang, Jeffrey Dean, Sanjay Ghemawat, Wilson C. Hsieh, Deborah A. Wallach, Michael Burrows, Tushar Chandra, Andrew Fikes, and Robert Gruber: "Bigtable: A Distributed Storage System for Structured Data", OSDI 2006, pp. 205-218

[12] Victor R. Lesser: "Encyclopedia of Computer Science", 4th edition. John Wiley and Sons Ltd. 2003, pp.1194–1196

[13] Ernesto Damiani, Sabrina De Capitani di Vimercati, Sushil Jajodia, Stefano Paraboschi, and Pierangela Samarati:"Balancing confidentiality and efficiency in untrusted relational DBMSs",ACM Conference on Computer and Comm. Security 2003, pp. 93-102

[14] Ernesto Damiani, Sabrina De Capitani di Vimercati, Mario Finetti, Stefano Paraboschi, Pierangela Samarati, and Sushil Jajodia: "Implementation of a Storage Mechanism for Untrusted DBMSs", IEEE Security in Storage Workshop 2003, pp. 38-46

[15] Sabrina De Capitani di Vimercati, Sara Foresti, Stefano Paraboschi, and Pierangela Samarati: "Privacy of outsourced data", In Alessandro Acquisti, Stefanos Gritzalis, Costos Lambrinoudakis, and Sabrina De Capitani di Vimercati: Digital Privacy: Theory, Technologies and Practices. Auerbach Publications (Taylor and Francis Group) 2007

[16] Valentina Ciriani, Sabrina De Capitani di Vimercati, Sara Foresti, Sushil Jajodia, Stefano Paraboschi, and Pierangela Samarati: "Fragmentation and Encryption to Enforce Privacy in Data Storage", ESORICS 2007, pp. 171-186

[17] Richard Brinkman, Jeroen Doumen, and Willem Jonker: "Using Secret Sharing for Searching", in Encrypted Data. Secure Data Management 2004, pp. 18-27

[18] Ping Lin and K. Selçuk Candan: "Secure and Privacy Preserving Outsourcing of Tree Structured Data", Secure Data Management 2004, pp. 1-17

[19] Valentina Ciriani, Sabrina De Capitani di Vimercati, Sara Foresti, Sushil Jajodia, Stefano Paraboschi, and Pierangela Samarati: "Combining fragmentation and encryption to protect privacy in data storage", ACM Trans. Inf. Syst. Secur. 13(3): (2010)

[20] Valentina Ciriani, Sabrina De Capitani di Vimercati, Sara Foresti, Sushil Jajodia, Stefano Paraboschi, and Pierangela Samarati: "Keep a Few: Outsourcing Data While Maintaining Confidentiality", ESORICS 2009, pp. 440-455

[21] Miseon Choi, Wonik Park, and Young-Kuk Kim: "A split synchronizing mobile transaction model", ICUIMC 2008, pp.196-201

[22] Henk C. A. van Tilborg: "Encyclopedia of Cryptography and Security", Springer 2005

[23] Ian T. Foster, Yong Zhao, Ioan Raicu, and Shiyong Lu: "Cloud Computing and Grid Computing 360-Degree Compared CoRR", abs/0901.0131: (2009)

[24] Hakan Hacigümüs, Balakrishna R. Iyer, Chen Li, and Sharad Mehrotra: "Executing SQL over encrypted data in the database-service-provider model", SIGMOD Conference 2002, pp. 216-227

[25] Dirk Düllmann, Wolfgang Hoschek, Francisco Javier Jaén-Martínez, Ben Segal, Heinz Stockinger, Kurt Stockinger, and Asad Samar: "Models for Replica Synchronisation and Consistency in a Data Grid", HPDC 2001, pp. 67-75

[26] Raju Halder, Shantanu Pal, and Agostino Cortesi: "Watermarking Techniques for Relational Databases: Survey, Classification and Comparison", in Journal of Universal Computer Science, vol. 16 (21), pp. 3164-3190

[27] H. Garcia-Molina and K. Salem, "Main Memory Database Systems: An Overview," IEEE Trans. Knowl. Data Eng. 4(6), 1992, pp. 509-516

[28] L. Bouganim and Y. Guo, "Database encryption," in Encyclopedia of Cryptography and Security, Springer, 2010, 2nd Edition

[29] E. Damiani, S. De Capitani di Vimercati, S. Foresti, S. Jajodia, S. Paraboschi, and P. Samarati, "Key management for multi-user encrypted databases," StorageSS, 2005, pp. 74-83

[30] D. Boneh and M. Hamburg, "Generalized Identity Based and Broadcast Encryption Schemes," ASIACRYPT, 2008, pp. 455-470

[31] V. Goyal, A. Jain, O. Pandey, and A. Sahai, "Bounded Ciphertext Policy Attribute Based Encryption," ICALP, 2008, pp. 579-591

[32] A. Fiat and M. Naor, "Broadcast Encryption," CRYPTO, 1993, pp. 480-491

[33] E. Damiani, S. De Capitani di Vimercati, S. Paraboschi, and P. Samarati, "Computing range queries on obfuscated data," IPMU, 2004

[34] C. Pu and A. Leff, "Replica Control in Distributed Systems: An Asynchronous Approach," SIGMOD, 1991, pp. 377-386

[35] http://en.wikipedia.org/wiki/In-memory_database. Retrieved 2011-07-22

[36] http://www.remote-dba.net/t_in_memory_cohesion_ssd.htm. Retrieved 2011-07-22

[37] www.hsqldb.org. Retrieved 2011-07-22